\documentstyle[psfig,nicv]{article}

\newcommand{\ctanb}{$^{13}$C($\alpha$,n)$^{16}$O~}

\newcommand{\ndb}{$^{142}$Nd~}

\newcommand{\gca}{Geoch. Cosmoch. Acta }

\newcommand{\prc}{Phys. Rev. C }
\newcommand{\prl}{Phys. Rev. Lett. }
\newcommand{\apjs}{ApJ Suppl. }

\begin{document}
%
%
\heading{%
$s$-process nucleosynthesis of $^{142}$Nd: \\
crisis of the classical model\\
%
}
\par\medskip\noindent
%
\author{Claudio Arlandini$^{1}$, Franz K{\"a}ppeler$^{1}$,
Klaus Wisshak$^{1}$, Roberto Gallino$^{2}$, 
Maria Lugaro$^{3}$, Maurizio Busso$^{4}$, 
Oscar Straniero$^{5}$}

\address{Forschungszentrum Karlsruhe, Institut f{\"u}r Kernphysik, 
          D-76021 Karlsruhe, Germany}
\address{Dipartimento di Fisica Generale, Universit{\`a} di Torino, 
          I-10125 Torino, Italy}
\address{Department of Mathematics, Monash University, Clayton, 
          Victoria 3168, Australia}
\address{Osservatorio Astronomico di Torino, I-10025 Torino, Italy}
\address{Osservatorio Astronomico di Collurania, I-64100 Teramo, Italy}

%
\begin{abstract}
The recently improved information on the stellar neutron capture
cross sections of neutron magic nuclei at $N =$ 82, and in particular
for
$^{142}$Nd, turned out to represent a sensitive test for models of
$s$-process nucleosynthesis. While these data were found to be
incompatible with the classical approach, they provide significantly
better agreement between the observed abundance distribution and the
predictions of models for low mass AGB stars.

\end{abstract}

\section{Introduction} 

In the last thirty years, $s$-process studies have been mainly
pursued either by nucleosynthesis computations in stellar models 
for the Thermally Pulsing Asymptotic Giant Branch (TP-AGB) phases of 
low and intermediate mass stars \cite{sp95,gp98} or by a 
phenomenological model, the so-called {\it classical approach},
developed with the intention to provide a possibility for a "model-free" 
description (see e.g. Ref.\cite{kap89}). 

Until recently, the two descriptions remained compatible within 
their respective uncertainties, while the \ctanb neutron source was 
recognized to play a major role on the AGB, and was assumed to occur 
in convective thermal pulses \cite{ir82,hi89,kap90}. Under such conditions, 
the classical analysis was considered suitable to extract 
"effective" conditions characterizing the stellar scenarios.

This situation  changed  drastically, when it was realized that 
$^{13}$C burns radiatively in the time interval between two successive 
He-shell instabilities \cite{sp95}. The interplay of the 
different thermal conditions for the $^{13}$C and $^{22}$Ne neutron 
sources became so complex as to be hardly represented by a single set 
of effective parameters, as commonly used by the classical approach 
(essentially neutron density $n_n$, temperature {\it T}, mean neutron 
exposure $\tau_0$). This is particularly true for the mathematical 
treatment of the  neutron irradiation, usually simplified through
an exponential distribution of exposures in phenomenological studies, 
$\rho$($\tau$) $\sim$ exp($-\tau$/$\tau_0$).  Stellar models now show that 
this distribution  
is definitely non-exponential. Any attempt to maintain the classical picture 
would require a larger number of free parameters, in contradiction
with  the basic reason for this approach as a model-independent guideline for stellar
calculations. In view of these conceptual differences, it is not 
surprising that the results of the classical analysis and of AGB models 
exhibit significant discrepancies. A region of the $s$-process path where this is
particularly evident, involves the neutron-magic nuclei at $N =$ 82,
including the s-only isotope $^{142}$Nd. The small cross sections 
of these nuclei act as bottle necks for the reaction flow and are,
therefore, sensitive to the characteristics of the neutron
exposure.

\section{The new $^{142}$Nd cross section and its implications for
$s$-process nucleosynthesis}

\begin{figure}
\centerline
{\vbox{\psfig{figure=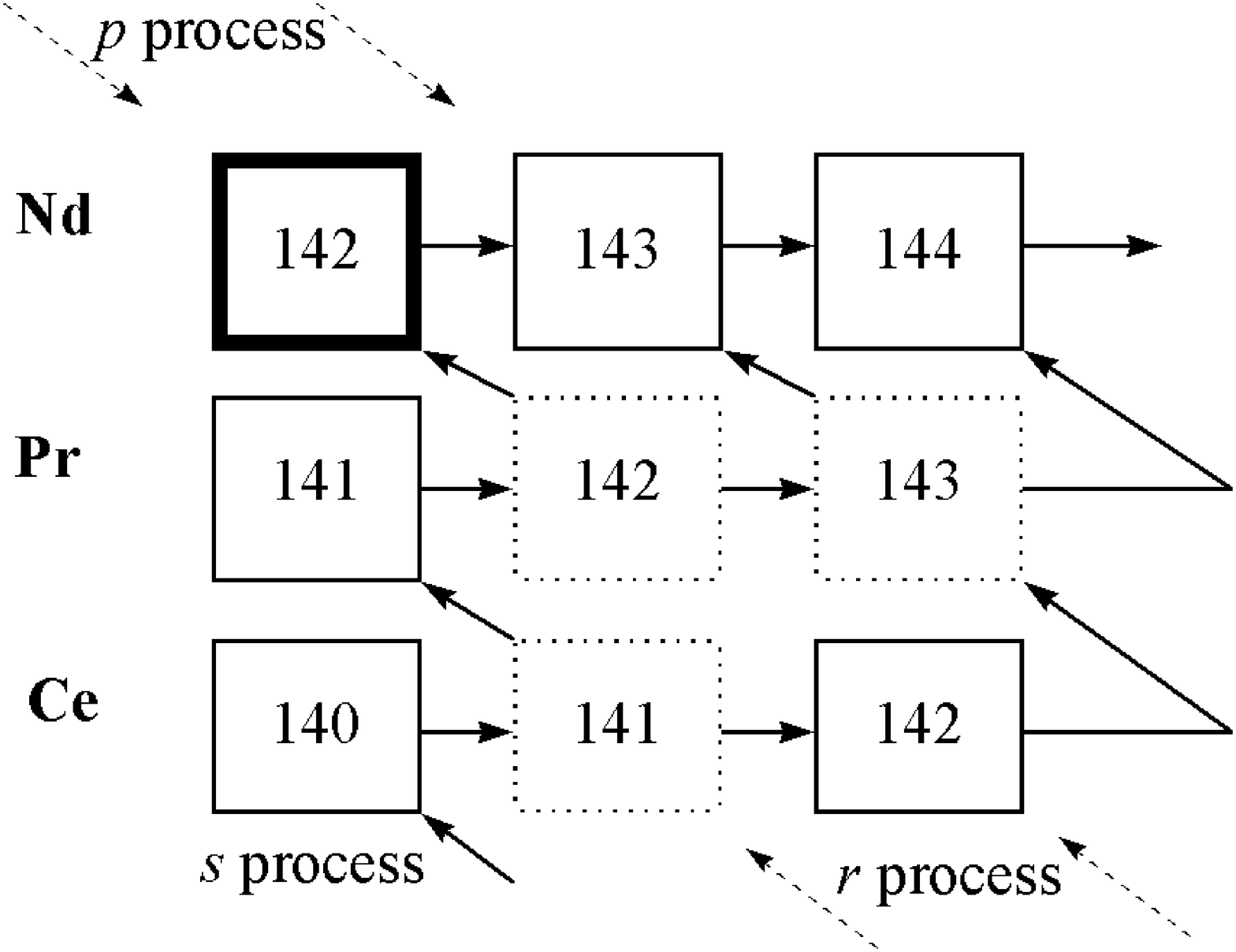,height=5.cm,width=7.5cm}}}
\caption[]{\small
The $s$-process path between Ce and Nd.}
\end{figure}

The $s$-process abundance of $^{142}$Nd is influenced 
by small branchings in the neutron capture path at $^{141}$Ce and 
$^{142}$Pr (Fig. 1).
The expected branching factor for bypassing $^{142}$Nd is about 5\%. 
While a meaningful analysis was long hampered by the 
uncertainties in the nuclear physics data, especially for \ndb,
accurate measurements of the
stellar (n,$\gamma$) cross sections of $^{140,142}$Ce \cite{kap96},
$^{141}$Pr \cite{vp98}, and of all stable Nd isotopes 
\cite{tp95,gup97,wip98a,wip98b} have been reported recently, along
with new calculations of the cross sections of the unstable branch 
point nuclei $^{141}$Ce and $^{142}$Pr \cite{kap96}. 
Significant discrepancies were found with the previously adopted 
cross sections, thus requiring an updated analysis.

\ndb is located immediately at the pronounced 
precipice of the $\left< \sigma \right>$$N_{s}$ curve, which is caused 
by the small (n,$\gamma$) cross sections at $N =$ 82. Hence, its cross 
section determines not only the branching analysis, but also the general
shape of the $s$-process distribution. Since the $\beta$-decay rates 
of $^{141}$Ce and $^{142}$Pr are almost independent of the stellar 
temperature \cite{ty87}, the branchings are completely defined, according to the classical model,
by the effective $s$-process neutron density, $n_n$, which is best
obtained from the neighboring branchings bypassing $^{148}$Sm:
$n_n = \left( 4.1\pm0.6 \right)\times 10^8$ cm$^{-3}$ \cite{tp95}.
This value together with a thermal energy 
of $kT =$ 30 keV and an electron density of $n_e = $ 5.4 
$\times$ 10$^{26}$ cm$^{-3}$ has been used for an $s$-process analysis 
with an updated reaction network.

\begin{figure}
\centerline
{\vbox{\psfig{figure=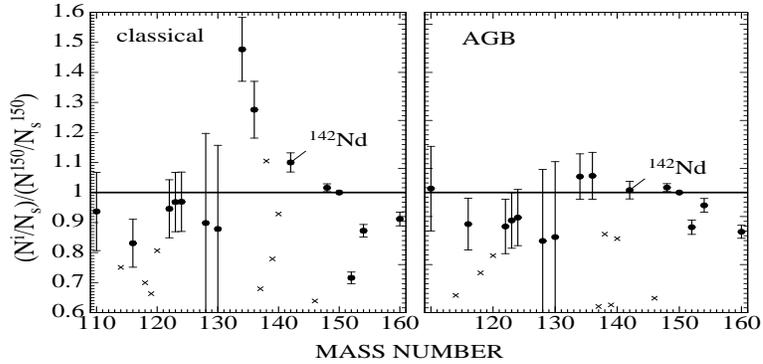,height=5.cm,width=10.cm}}}
\caption[]{\small
Overproduction factors for $s$-only nuclei normalized to $^{150}$Sm for the best 
reproduction of the solar distribution 
obtained with the classical model ({\it left}) and with the 
stellar model for a 2 $M_{\odot}$ star ({\it
right}). Full circles indicate $s$-only nuclei.}
\end{figure}

The best fit to the solar system distribution for the $s$-nuclei
belonging to the main component is obtained for a mean neutron
exposure $\tau_0 = $(0.296$\pm$0.003) [$kT$/30]$^{1/2}$ mbarn$^{-1}$.
The left panel of Fig. 2 provides evidence for a significant overproduction of
$^{142}$Nd with respect to the reference isotope $^{150}$Sm, 
despite part of the reaction flow bypasses this isotope.
The overproduction exceeds by far the uncertainties
of the cross section (2\% at 30 keV) and of the solar abundance, 
the abundance ratio between the chemically related rare earth 
elements Nd and Sm being known to 1.8\% \cite{ag89}. Furthermore, the 
overproduction may be even larger in case of
a non-neglibible $p$-process contribution.
Therefore, $^{142}$Nd provides the first significant hint that the simple 
assumptions of the classical approach are not adequate to describe 
the $s$-process near magic neutron numbers. Failures of the 
classical model have been found previously for the $s$-only nuclei 
$^{116}$Sn and $^{134,136}$Ba \cite{vp94,wip96}, but these discrepancies 
were masked by the large uncertainties of the solar 
Sn and Ba abundances. 

A description of the stellar model and the adopted reaction 
network can be found in Ref.\cite{gp98}. Updated $s$-process 
calculations have been made using stellar evolutionary computations 
with the FRANEC code \cite{cs89} for low mass stars, from the main 
sequence
up to end of the AGB phase. A relatively large range of masses 
(1.5 $\leq$ $M/M_{\odot} \leq$ 3) and metallicities (-0.4 
$\leq$ [Fe/H] $\leq$ 0) have been examined as well as  
the influence of various mass-loss rates.
 
The best reproduction of the solar distribution of the main
$s$-component was obtained for a 2 $M_{\odot}$ star with $Z = 1/2$ 
$Z_{\odot}$, and a mass loss rate according to Reimers with 
$\eta = 0.75$. 

The general improvement with respect to the classical solution 
(Fig. 2), is striking, especially since no fitting 
procedure was applied. $^{134,136}$Ba are now overproduced by only 
5\%, a value compatible with the uncertainties of the neutron capture 
cross sections and solar abundances, thus avoiding the 20\% 
correction of the solar barium abundance suggested by the 
classical analysis \cite{vp94}. Similarly, also $^{116}$Sn,
that had required a 15\% correction of the solar tin abundance
\cite{wip96}, is now reproduced within uncertainties.
As for $^{142}$Nd, the new cross section improves the 
situation drastically, resulting in very good agreement
with the solar abundance. The previously overestimated 
$^{142}$Nd cross section had always led to a persisting 
30\% deficiency of the $^{142}$Nd abundance, incompatible with the 
expected $p$-process contribution of only a few percent
\cite{PHR90}.

Actually, to reproduce the $s$-isotopes in the solar system requires 
a more complex analysis including the chemical evolution of the Galaxy. Also in this context, 
the updated (n,$\gamma$) cross sections yield a satisfactory
description of the solar $^{142}$Nd abundance \cite{t+98,g+98}.

\section{$r$-process residuals}

\begin{figure}
\centerline
{\vbox{\psfig{figure=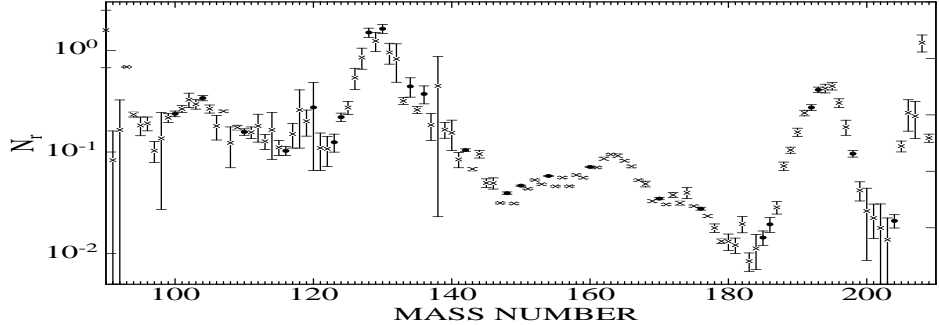,height=4.7cm,width=\textwidth}}}
\caption[]{\small
The $r$-process residuals $N_r = N_{\odot} - N_s$ obtained with
the $s$ abundances from the stellar model (see text).}
\end{figure}
The $r$-process residuals $N_r = N_{\odot} - N_s$ in Fig. 3 were 
calculated with the $s$-abundance average from the models for 1.5 
and 3 $M_{\odot}$ stars, which represents a good approximation
of the distribution obtained with a chemical evolution study.
Compared to the $r$ residuals derived from the classical $s$ abundances
the plot of Fig. 3 shows only minor differences. However,
it should be noted that the stellar models yield a 20\% $r$-process component
of the solar barium abundance, in excellent 
agreement with recent observations in low metallicity stars \cite{gvol}.    
\vfill
\end{document}